# A Safe Regression Testing Technique for Web Services based on WSDL Specification

Tehreem Masood[1], Aamer Nadeem[1], Gang-soo Lee[2]

[1]Center for Software Dependability,
Mohammad Ali Jinnah University (MAJU), Islamabad, Pakistan
tehreem_maju@yahoo.com , anadeem@jinnah.edu.pk

[2]Department of Computer Engineering,
Hannam University, Korea
gslee@hannam.ac.kr

**Abstract.** Specification-based regression testing of web services is an important activity which verifies the quality of web services. A major problem in web services is that only provider has the source code and both user and broker only have the XML based specification. So from the perspective of user and broker, specification based regression testing of web services is needed. The existing techniques are code based. Due to the dynamic behavior of web services, web services undergo maintenance and evolution process rapidly. Retesting of web services is required in order to verify the impact of changes. In this paper, we present an automated safe specification based regression testing approach that uses original and modified WSDL specifications for change identification. All the relevant test cases are selected as reusable hence our regression test selection approach is safe.

## 1 Introduction

Web services have become center of attention during the past few years. It is a software system designed to support interoperable interaction between different applications and different platforms. A system in which web services are used is named as web services based system. Web services use standards such as Hypertext Transfer Protocol (HTTP), Simple Object Access Protocol (SOAP) [13], Universal Description, Discovery, and Integration (UDDI), Web Services Description Language (WSDL) and Extensible Markup Language (XML) [3] for communication between web services through internet [1].

Maintenance is the most cost and time consuming phase of software life cycle, it requires enhancement of previous version of software to deal with the new requirements or problems. As modifying software may incur faults to the old software, testing is required. It is very difficult for a programmer to find out the changes in software manually, this is done by making comparison of both previous test results and current test results being run. Now the changed or modified software needs testing known as regression testing [2].

Regression testing is performed during and after the maintenance to ensure that the software as a whole is working correctly after changes have been made to it. Basic regression testing steps includes change identification in modified version of the system, impact of changes on other parts of the system, compatibility of both changed part and indirectly affected part with the baseline test suite, removing invalid test cases and selecting a subset of baseline test suite that is used for regression testing [2].

Significant research has been carried out on testing of web services [12] but there is limited amount of work on regression testing of web services. Most of the existing approaches for regression testing of web services are code based but no work is available on specification based regression testing of web services.

In web services, only web service provider has the source code and both web service broker and user only have the specification. Provider is not willing to share the source code [1]. So from the perspective of broker and user, specification based regression testing is needed. A change may occur in web service functionality or behavior with no interface change, specification will not change. But if a change occurs in interface, specification will also be changed [6]. Our focus is interface change. Further details about changes are explained in section III.

WSDL plays very important role in web services. It is an XML document used to describe web services. It has four major elements that are Types, Messages, PortType and Binding [8]. The main concern of our approach is Type element of WSDL specification [8]. WSDL specification uses an XML Schema [14], which is used to define types used by web service. XML schema defines simple types and complex types [14]. For simplicity, we will only consider higher level complex types. Complex type within a complex type is not considered because the depth of the tree increases.

We have applied boundary value analysis [10] on data type level changes and selected reusable test cases [11]. Test suite classification of Leung and white [11] is used in this paper. The proposed approach selects all the relevant test cases as reusable test cases which is explained by the help of an example. Safety is defined as all the relevant test cases are used [2].

The remaining paper is organized as follows: Section II includes related work in the area of regression testing of web services. Section III discusses the proposed approach for selective regression testing. In the end conclusion of the paper is presented in Section IV.

## 2 Related Work

Ruth, *et al*. [4] presented an approach to apply a safe regression test selection technique to Java web services. Their approach is based on Java-based control flow graph named as Java Interclass Graph (JIG).They have created JIG by performing static and dynamic analysis of code. They identified dangerous edges by comparing old and new JIG. Then they compared the table of edges covered by the tests with the set of dangerous edges to identify the tests to be performed. They provided a simulation tool.

Ruth, *et al*. [5] presented a framework to apply a safe regression test selection technique to generic web services. There technique is based on control flow graph for service involved in the regression testing activity. The idea is that Control Flow Graphs (CFG) should be able to highlight the changes that can cause regression testing. They also discussed that publishing test cases is useful.

Penta, *et al*. [6] used test cases as a contract between service provider and system integrator. They considered dynamicity as an important characteristic of service-based applications, and performed online tests, i.e., to execute tests during the operation phase of the service-based application. They discussed their approach with respect to some scenarios and used some QoS assertions for performing service regression testing. They didn't focus the changes in the specifications. They provided a toolkit for generating XML-encoded test suite.

Khan and Heckel [7] presented a model-based approach for regression testing of web services. They identified changes and impact of changes by using models that are used to describe the service interface. For external behavior they used finite state automatons and for data dependencies, bipartite dependency graph is used where the nodes represents methods and classes. Then a method for test case selection is presented.

## 3  Proposed Approach

In web services, a change may occur in web service functionality or behavior and interface is not changed, in this case specification will not change and old test cases can be used. But if a change occurs in interface, specification will also be changed. In this case, some required old test cases can be selected and there is a need to develop some new test cases for regression testing [6]. The proposed approach focuses on interface change.

A WSDL specification has four major elements which are messages, types, binding and port type [8]. A message provides an abstract definition of the data which is being transmitted. A binding is used to define format of message and protocol details for operations and messages. A port type represents a set of abstract operations. A type is used to provide a data type definition, used to describe the exchanged message which is then used by a web service. WSDL specification uses an XML Schema which is used to define types used by web service [14].

Figure 1 shows the overall architecture of our proposed approach for specification based regression testing of web services. The major components of our approach are parser, comparator and regression test selector. Original WSDL specification of web service is named as baseline WSDL, when initially a web service is build. Modified WSDL specification is named as delta WSDL, when a web service is changed. Major components are explained below.

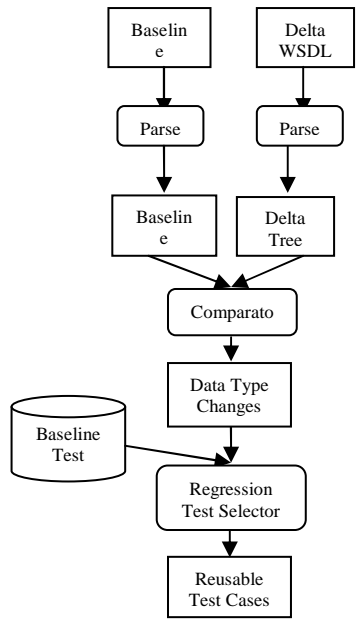

**Fig. 1.** Abstract Model of the proposed approach

```
Input: Wsdl specification    output: baseline and
delta tree
Step 1:
Select type element of WSDL specification.
Step 2:
Select element to be modeled as root node. Select
name of element as its attribute along with its
value enclosed with an equal sign.
Step 3:
If complex type of the root element exists then
an outgoing directed edge is formed from root
node which is connected to a complex type node
having name as its attribute along with its value
enclosed with an equal sign.
Step 4:
For each sub element of complex type, a new node
is generated, having element as name of the node
and a number of the element.
Associate all attributes and their values with
their respective nodes, in the same sequence as
described in the schema.
Step 5:
Repeat step 3 for all the sub elements of complex
type
```
**Fig. 2.** Algorithm for generating tree of datatype

### 3.1 Parser

As described earlier, the main concern of our approach is Type element of WSDL specification. It provides a data type definition for describing messages exchanged.

WSDL uses an XML schema to describe types used by Web service [14]. XML Schema defines simple types and complex types. Simple types include either built in primitive data types or derived data types and data type facets. Built in primitive types include string, float, etc. Derived data types include integer, etc. Complex type is any user defined data type. A facet is a constraint which is used to set the values for a simple type like length, minInclusive, etc [14]. Here we are taking facets as attributes of sub elements. Parser takes original and modified WSDL specifications as input and generates tree for type element of the WSDL specification. An algorithm for generating trees for both original and changed WSDL specifications is given in Fig 2. If any attribute of element, complex type and sub element have no value specified in XML schema then the attribute value is considered as null.

**Example: MortgageIndex**

MortgageIndex is a Web service used to provide monthly, weekly and Historical Mortgage Indexes. There are many possible Adjustable Rate Mortgage (ARM) indexes. Some common mortgage indexes are 12-Month Treasury Average (MTA), Treasury bill (T-Bill), etc [9]. For example if a borrower thinks that interest rates are going to rise in the future, the T-Bill index would be a more economical choice than the one-month LIBOR index because the moving average calculation of the T-Bill index creates a lag effect.

This web service has four basic operations, i.e., GetCurrentMortgageIndexByWeek, GetCurrentMortgageIndexMonthly, GetMortgageIndexByMonth, GetMortgageIndexByWeek. Here we are taking one operation for explanation which is GetMortgageIndexByMonth. This operation takes month and year as input and provides ARM indexes for the specified values. Both month and year are of type int [9]. XML schema for this operation is provided in Fig 3.

```
    Example: XML Schema
<s:element name="GetMortgageIndexByMonth">
<s:complexType>
<s:sequence>
<s:element maxOccurs="1" minOccurs="1" name="Month" type="s:int"  maxInclusive ="12" minInclusive ="1" />
<s:element maxOccurs="1" minOccurs="1" name="Year" type="s:int" maxInclusive ="2007"  minInclusive ="1990"/>
</s:sequence>
</s:complexType>
</s:element>
<s:element  name="GetMortgageIndexByMonthResponse">
<s:complexType>
<s:sequence>
<s:element minOccurs="1" maxOccurs="1" name="GetMortgageIndexByMonthResult" type="tns:MonthlyIndex"/>  </s:sequence>
</s:complexType>
</s:element>
```

**Fig. 3.** Original XML Schema for Element GetMortgageIndexByMonth [9]

A runtime view of the operation GetMortgageIndexByMonth is shown in Fig 4.

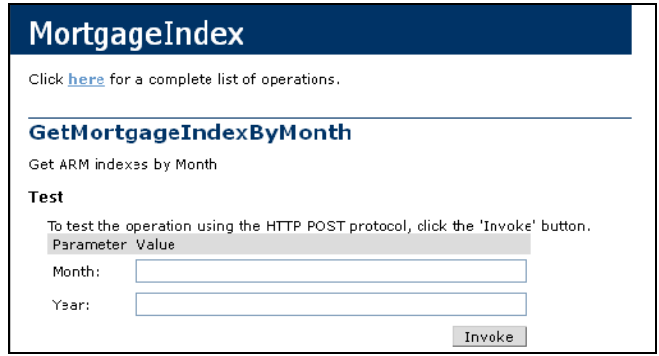

**Fig 4.** A runtime view of GetMortgageIndexByMonth

The resulting baseline tree generated from the original schema is shown in Fig 5.

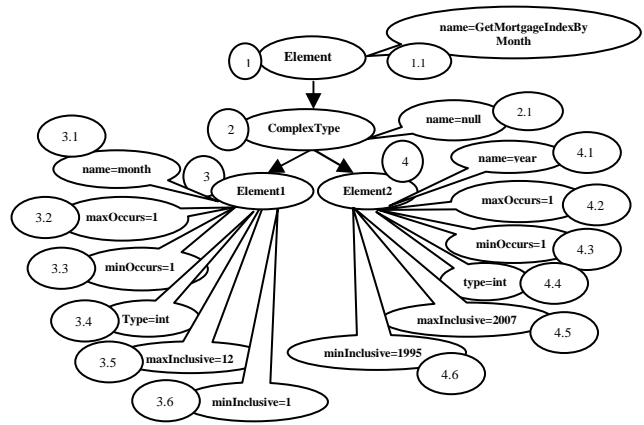

**Fig. 5.** Baseline tree T for complex datatype

Baseline tree for complex Datatype of Fig 3 is shown in Fig 5. In Fig 5 an oval shape represents a node e.g. here Element, ComplexType, Element1 and Element2 are nodes. An oval callout represents attributes and facets of simple type e.g. name, minOccurs, maxOccurs and type are attributes and minInclusive, maxInclusive are facets of simple type int. Here facets are also considered as attributes. The procedure for generating tree from schema as shown in Fig 3 is explained below.

In this example, first parser takes element of the type and generates a tree for it. A node shape is drawn and named it as Element. Then an attribute shape is attached with this element and enclosed the name as an attribute and a value of this attribute in it. For example here attribute is name and value is GetMortgageIndexByMonth. So Name= GetMortgageIndexByMonth is written inside the attribute shape. Then the next element is complex type, a node shape is drawn and named it as complex type. Then draws an outgoing directed edge from the root node and connect it to the new node. Then an attribute shape is attached with this new node and enclosed the name as its attribute and value of this attribute in it. Here a null value is assigned to the attribute name of ComplexType as it has no name specified in Fig 3. In Fig 3 there are two sub elements of ComplexType. First sub element is month. Now as its type is int which is a primitive data type, a new node is drawn and named it as Element1. Then an outgoing directed edge is drawn from ComplexType node to this new node. Then attribute shapes are attached with this new node for every attribute of Element 1 and enclosed the name and values of these attributes one by one. For example here 1$^{st}$ attribute is minOccurs having value 1, so in the attribute shape minOccurs =1 is written. Similarly all other attributes are drawn. Repeat the same procedure for the all other sub elements of ComplexType. For naming scheme circle having 1value is attached with the root node, as it is the first node of the tree. 1.1 is attached with the attribute name of root node. 2 is attached with the 2$^{nd}$ node named as complex type. 2.1 is attached with the attribute name of complex type node. Same is the case with other nodes and attributes. The resulting baseline tree model for element GetMortgageIndexByMonth generated by applying the above steps is shown in Fig 5. Now suppose a change occurs in the attributes of element2 (year) described in Fig 3. The changed schema is described in Fig 6. Here the values of the attributes minInclusive and maxInclusive are changed to 1995 and 2000 respectively. Remaining schema is same. Then the tree is generated from the modified schema by applying the same steps presented in Fig 2. As the only change is in the values of minInclusive and maxInclusive of element2 (year), so only these attribute values are changed in the resulting delta tree.

```
    Example: XML Schema
<s:element name="GetMortgageIndexByMonth">
<s:complexType>
<s:sequence>
<s:element maxOccurs="1" minOccurs="1" name="Month" type="s:int"  maxInclusive ="12"  minInclusive ="1" />
<s:element maxOccurs="1" minOccurs="1" name="Year" type="s:int" maxInclusive ="2000"  minInclusive ="1995"/>
</s:sequence>
</s:complexType>
</s:element>
<s:element  name="GetMortgageIndexByMonthResponse">
<s:complexType>
<s:sequence>
<s:element minOccurs="1" maxOccurs="1" name="GetMortgageIndexByMonthResult" type="tns:MonthlyIndex" />
</s:sequence>
</s:complexType>
</s:element>
```

**Fig. 6.** Modified XML schema

The resulting delta tree generated from the modified schema is shown in Fig 7.

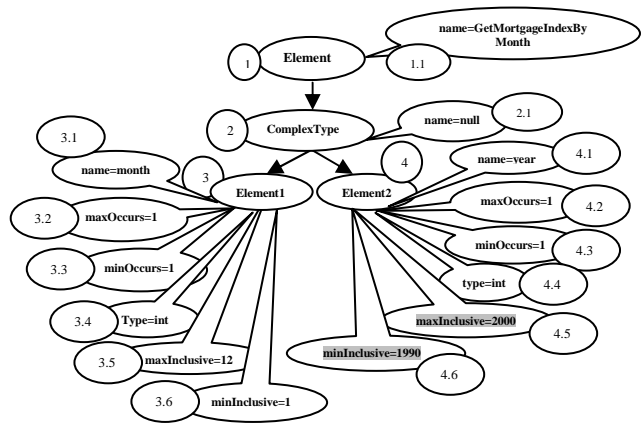

**Fig. 7.** Delta tree T' for complex datatype

### 3.2 Comparator

We define data type changes in our approach. Type element of the WSDL specification provides a data type definition for describing messages exchanged. In Fig 8 e is used for representing the root element of complex type in baseline tree and e' is used for representing the root element of complex type in delta tree. c is used to represent complex type in baseline tree and c' is used to represent complex type in delta tree. ei is the instance of sub element of complex type in baseline tree and ej is the instance of sub element of complex type in delta tree. $1^{st}$ the root element attribute name and value of Fig 5 and Fig 7 are compared, as here they are same so step 2 is executed. In step 2 name and value of the complex type attribute are compared, as they are also same, so we execute step 3 and 5.

In step 3 we check the sub elements of complex type one by one, $1^{st}$ element of complex type node of Fig 5 and Fig 7 when compared, they are same, and so step 4 is executed for its every attribute. In step 4 all attribute names and values are compared one by one, they are same so check if there is any other element of the complex type. As there is another sub element denoted by element2, so the name and value of element2's attribute of both Fig 5 and Fig 7 are compared, as they are same so step 4 is executed for its every attribute. In step 4 all the attribute names and values are compared one by one, in this case two values of attributes maxInclusive and minInclusive are different so this change is detected by comparator shown in Table 1. $1^{st}$ is the change in the attribute maxInclusive value and $2^{nd}$ is the change in the attribute minInclusive of Element 2.

```
Input: baseline and delta tree Output: changes
Variables: e denotes root node of baseline tree
e'denotes root node of delta tree
c denotes complex type node of baseline tree
c' denotes complex type node of delta tree
ei denotes instance of sub element of complex type in
baseline tree
ej denotes instance of sub element of complex type in
delta tree.
Step 1:
Compare e.name and e'.name. If matched then If
e.value==e'.value then execute step 2 else root node
deleted from baseline tree. If e.name! =e'.name and
e.value==e'.value then content of the attribute
changed. Execute step 2. Else If e.name! =e'.name and
e.value! =e'.value then root node deleted.
Step 2:
Compare c.name and c'.name. If matched then If
c.value==c'.value then execute step 3, 5 else complex
type node deleted from baseline tree. If c.name!
=c'.name and c.value==c'.value then content of the
attribute changed. Execute step 3, 5. Else If c.name!
=c'.name and c.value! =c'.value then complex type node
deleted.
Step 3: Check_subElement (c, c')
   For each child ei of c
   For each child ej of c'
Compare ei.name and ej.name. If matched then If
ei.value==ej.value then execute step 4 for every
attribute else sub element node deleted from baseline
tree. Execute step 4. If ei.name! =ej.name and
ei.value==ej.value then content of the attribute
changed. Execute step 4. Else If ei.name! =ej.name and
ei.value! =ej.value then attribute deleted from
baseline tree. Execute step 4 for every attribute.
Step 4:
Compare attribute name and value. If matched then
repeat step 4 for other attributes Else attribute
changed. Repeat step 4 for other attributes.
Step 5: Check_subElementAdded (c, c')
   For each child ej of c
   For each child ei of c'
   If ej.name==ei.name then If ej.value==ei.value then
matched else sub element added in delta tree. If
ei.name! =ej.name and ei.value==ej.value then content
of the attribute changed. Else If ei.name! =ej.name
and ei.value! =ej.value then attribute added in delta
tree.
```

**Fig. 8.** Change detection algorithm

The detected changes are shown below in Table 1.

**Table 1:** Detected changes

| |
|---|
| 4.5 changed |
| 4.6 changed |

### 3.3 Regression Test Selector

Finally regression test selector takes baseline test suite and data type changes as input, and categorizes the test suite. Baseline test suite is categorized into obsolete and reusable test cases [11]. Obsolete test cases are those test cases that are invalid for the delta version. They are invalid because the elements may be changed or deleted from the baseline version. Reusable test cases are those test cases that are still valid for the delta version after applying boundary value conditions. We perform boundary value analysis for test case selection [10]. Criteria that we are using for boundary value analysis is max value, min value, max-1, min+1 and 1 random value which should be greater than min value and less then max value. For example by applying these conditions on specification shown in Fig 3, we get for month: 1,2,11,12,7 and for year: 1990, 1991, 2006, 2007 and 1998. First of all baseline test suite for original specification tree T is shown in Table 2 by applying the above boundary value conditions. Every value of month is combined with every value of year.

**Table 2:** Baseline test suite

| TC1=1,1990 | TC10= 7,1991 | TC19= 12,2007 |
|---|---|---|
| TC2=2,1990 | TC11=1,2006 | TC20= 7,2007 |
| TC3=11,1990 | TC12=2,2006 | TC21= 1,1998 |
| TC4=12,1990 | TC13= 11,2006 | TC22= 2,1998 |
| TC5= 7,1990 | TC14= 12,2006 | TC23= 11,1998 |
| TC6=1,1991 | TC15= 7,2006 | TC24= 12,1998 |
| TC7=2,1991 | TC16= 1,2007 | TC25= 7,1998 |
| TC8= 11,1991 | TC17= 2,2007 | |
| TC9=12,1991 | TC18= 11,2007 | |

Algorithm of regression test selector is shown in Fig 9. Now for the modified tree T' in Fig 7, again boundary value analysis is performed for checking the usability of baseline test suite. The resulting reusable test cases are shown in Table 3.

```
Step 1:
Perform boundary value analysis on the boundaries of
delta version
Step 2:
Compare new boundaries with the baseline boundaries
Step 3:
Select test cases from the baseline test suite that
are still valid for delta version named as reusable
test cases.
Step 4:
Discard the obsolete test cases from the baseline test
suite that are no longer valid for the delta version.
```

**Fig. 9.** Test case selection algorithm

Now the boundary values for year become 1995, 1996, 2000, 1999, 1997. The test cases that are still valid from the baseline test suite after applying the changed boundary value conditions are shown in Table 3 which is known as reusable test cases.

**Table 3:** Reusable test cases

| |
|---|
| TC21=1,1998 |
| TC22=2,1998 |
| TC23= 11,1998 |
| TC24=12,1998 |
| TC25= 7,1998 |

A regression test suite is considered as safe, if it includes all the test cases covering the whole changed part of the system as well as the whole indirectly affected part of the system. A safe regression test suite can have other test cases from the baseline test suite that are covering the unchanged part of the system. Here all the relevant test cases are used as reusable test cases. Hence our test case selection approach is safe.

**Case 1: Attribute changed**
If values of any attribute change then there can be impact on test cases. Check the type attribute of sub element and min and max inclusive or any other range attribute if there. If value of attribute name of root element is changed then it means old element is deleted. Check other attributes as well. If value of attribute name of complex type is changed then it means old complex type is deleted. Check other attributes as well. If value of all attributes of all sub elements. E.g. name, type, minOccurs, maxOccurs, minInclusive, maxInclusive etc is changed then it means value is changed. If maxInclusive and minInclusive values are changed then test cases will be selected from the baseline test suite according to the new values of the minInclusive and maxInclusive. If type value is changed then check the compatibility of new type with the previous one.

**Case 2: Types:** Check the compatibility of old and new types.
e.g. If int is changed to float and minInclusive and maxInclusive values are same then test cases will be selected according to min and max inclusive values but if there is no max and min inclusive values then test cases will not be selected from baseline test suite. If float is changed to int and minInclusive and maxInclusive values are same then test cases will be selected according to min and max inclusive values but if there is no max and min inclusive values then test cases will not be selected from baseline test suite. If int is changed to string, then test cases will not be selected from baseline test suite. If string is changed to int, then test cases will not be selected from baseline test suite. Same is the case with other types.

**Case 3: Node**
**Deleted:** If any node is deleted, then all its attributes are also deleted. If complex type node is deleted, then all its attributes are also deleted and all baseline test cases will be removed

**Added:** If any node is added, then check its attributes and values. Select test case according to the new values.

## 4  Conclusion

In this paper, we presented a specification based regression test selection approach for web services based on WSDL specification of web service. The proposed approach is a safe regression testing technique as it selected all test cases which exercise the modified parts of web service. A proof-of-concept tool has also been developed to support our approach.

## Ackmowledgement

This work was supported by the Security Engineering Research Center, under research grant from the Korean Ministry of Knowledge Economy.